\documentclass[superscriptaddress,twocolumn,prl]{revtex4-2}
\usepackage[utf8]{inputenc}
\usepackage{amsmath}
\usepackage{braket}
\usepackage{graphicx}
\usepackage{amsfonts}
\usepackage{pgfplots}
\usepackage{csquotes}
\usepackage{hhline}
\usepackage{amssymb}
\usepackage{listings}
\usepackage{color}
\usepackage{ulem}
\usepackage{booktabs}
\usepackage{comment}
\usepackage{lipsum}
\usepackage{dcolumn}
\usepackage{enumitem}
\usepackage{tabularx}
\usepackage{float}
\usepackage{caption} 
\usepackage{subcaption}
\makeatletter
\newcommand{\ssymbol}[1]{^{\@fnsymbol{#1}}}
\makeatother
\definecolor{codegreen}{rgb}{0,0.6,0}
\definecolor{codegray}{rgb}{0.5,0.5,0.5}
\definecolor{codepurple}{rgb}{0.58,0,0.82}
\definecolor{backcolour}{rgb}{0.95,0.95,0.92}

\newcommand{\kc}{k_c}

\newcommand{\OTF}{\mathrm{OTF}}
\newcommand{\Tr}{\mathrm{Tr}}
\newcommand{\TrAbs}{\mathrm{TrAbs}}

\newcommand{\psicap}{\psi}
\usepackage{mathtools}
\DeclarePairedDelimiter\abs{\lvert}{\rvert}

\lstdefinestyle{mystyle}{
    backgroundcolor=\color{backcolour},   
    commentstyle=\color{codegreen},
    keywordstyle=\color{magenta},
    numberstyle=\tiny\color{codegray},
    stringstyle=\color{codepurple},
    basicstyle=\footnotesize,
    breakatwhitespace=false,         
    breaklines=true,                 
    captionpos=b,                    
    keepspaces=true,                 
    numbers=left,                    
    numbersep=5pt,                  
    showspaces=false,                
    showstringspaces=false,
    showtabs=false,                  
    tabsize=2
}

\newcommand{\cNH}{c_{\mathrm{NH}}}
\newcommand{\cH}{c_{\mathrm{H}}}
\newcommand{\cQ}{c_{\mathrm{Q}}}

\usepackage[colorlinks=true,linkcolor=blue,citecolor=blue,urlcolor=blue]{hyperref}

\newcolumntype{C}{>{\centering\arraybackslash}X}

\begin{document}

\author{Aakash Warke}
\email{aakash.warke@physics.ox.ac.uk}
\affiliation{Clarendon Laboratory, University of Oxford, Parks Road, OX1 3PU, Oxford, United Kingdom}

\author{Aonan Zhang}
\affiliation{Clarendon Laboratory, University of Oxford, Parks Road, OX1 3PU, Oxford, United Kingdom}

\author{A. I. Lvovsky}
\affiliation{Clarendon Laboratory, University of Oxford, Parks Road, OX1 3PU, Oxford, United Kingdom} 


\title{Quantum-limited imaging using diffractive optical neural networks}

\begin{abstract}
We cast general imaging as multiparameter quantum estimation of band-limited spatial-frequency amplitudes. For separable (single-copy) measurements, we compute precision limits using semidefinite programming to evaluate the Nagaoka–Hayashi Cramér–Rao bound. We then introduce an architecture for a measurement apparatus based on diffractive optical neural networks and photon counting that saturates this bound. Extending the framework to arbitrary objects and many amplitudes, we show image reconstructions in which our architecture recovers fine features at the quantum limit, outperforming direct imaging. Together, these results open a scalable route to saturating multiparameter quantum limits in superresolution microscopy, telescopy, and remote sensing.
\end{abstract}

\begin{keywords}{Quantum imaging, superresolution microscopy, fluorescent microscopy, linear optics}\end{keywords}

\maketitle
\captionsetup{justification=raggedright,singlelinecheck=false}

\textit{Introduction.---\label{WZL_SectionI}} Optical imaging aims to reconstruct arbitrary objects faithfully despite limitations imposed by the available optics. Diffraction on the objective lens has been known for centuries to be the primary limitation on the resolution of far-field  imaging. In the photon-starved regime, the diffractive loss of resolution is exacerbated by the shot noise associated with the quantum nature of light, which affects high spatial frequency components of the object particularly strongly. 

To address imaging under these conditions, we consider the optical field collected by the imaging system to be a quantum state, with the imaging apparatus performing a measurement on that state. From this perspective, imaging is fundamentally a quantum sensing problem, and the ultimate limits of imaging resolution are determined by quantum mechanics. This viewpoint has already transformed several canonical tasks related to imaging. By an appropriate measurement, the separation between two incoherent point sources can be estimated precisely, even if that separation is below the diffraction limit ~\cite{Tsang2016,Paur2016,Tham2017,Boucher2020,Lvovsky2026}. Similar advances have followed for estimating centroids and moments of subdiffraction objects~\cite{Tsang2017moments,Zhou2019,Bisketzi2019,Tsang2019semiparametric,Tan2023}. Extending this framework from estimating a few parameters to arbitrary imaging, however, has remained \textit{terra incognita}. An image is described by a continuum of degrees of freedom, making it unclear how to even formulate imaging as a quantum estimation problem. 

The obstacle is not only the scaling of parameters, but also the complexity associated with quantum multiparameter estimation. Every single parameter carries its own optimal observable, the symmetric logarithmic derivative (SLD) of the quantum state. This leads to the fundamental precision limit set by the quantum Cramér–Rao bound (QCRB), which is the inverse of the quantum Fisher information matrix (FIM)~\cite{Helstrom1969}. But when multiple parameters need to be estimated, such that their corresponding SLDs fail to commute, no measurement can reach the QCRB for all parameters at the same time. A receiver must compromise between them, and this compromise costs precision~\cite{Szczykulska2016,Albarelli2020perspective,Liu2020,Haidong2026}, giving rise to additional precision benchmarks. If no limitations on the measurement are imposed, the quantum precision limit is the Holevo Cram\'er-Rao bound~\cite{Holevo1982,Albarelli2019, Almeida2025},  attainable asymptotically by means of collective (entangling) measurements across infinitely many copies of the quantum state. Although some ideas for such collective measurements --- such as storage in a photonic quantum memory --- have been proposed~\cite{Haidong2026}, they are hard to realize in experiment. In contrast, separable or single-copy measurements act independently on each photon and are far more readily implemented. For this class of measurements, the tight limit is set by the Nagaoka--Hayashi Cramér-Rao bound (NHCRB)~\cite{Nagaoka1989,Hayashi1999}, which was recently shown to admit efficient computation as a semidefinite program (SDP)~\cite{Conlon2021}. To summarize, the following inequality holds in general:
\begin{equation}
\Tr[\text{Cov}(\hat{\boldsymbol{\theta}})]\;\ge\;\Tr[F^{-1}]
\;\ge\;\cNH\;\ge\;\cH \ge\;\cQ,
\label{WZL_Eq3}
\end{equation}
where $\text{Cov}(\hat{\boldsymbol{\theta}})$ is the covariance matrix of an unbiased estimator $\hat{\boldsymbol{\theta}}$, $F$ is the FIM of the corresponding measurement, $\cNH$ the NHCRB, $\cH$ the Holevo bound, and $\cQ=\Tr[Q^{-1}]$ the QCRB ($Q$ is the quantum FIM). It has not yet been studied where the latter three limits lie for general imaging and how they compare to each other.


Furthermore, it is not known how to construct an apparatus that would achieve the relevant quantum bound. A straightforward approach to this problem would be in two steps: first, to find the optimal basis and second, to design the hardware that implements the measurement in this basis. For example, for estimating two point source separation, measuring the collected field in the basis of Hermite-Gaussian modes saturates the QCRB~\cite{Tsang2016,Paur2016,Boucher2020}; multi-plane light conversion enables mode sorting into this basis, albeit with significant limitations~\cite{Morizur2010,Fontaine2019,Zhang2023}. 

In this Letter, we make three principal contributions. First, we formulate the quantum estimation problem of general imaging. We do this by introducing a natural finite-dimensional parametrization of an arbitrary 1D or 2D object through the amplitudes of the band-limited spatial-frequency modes transmitted by the imaging system. Second, we establish the corresponding quantum precision limits for single-copy measurements by evaluating the NHCRB through semidefinite programming. We find that this limit separates from the QCRB as soon as the number of parameters exceeds one, and that the separation grows steadily as more parameters are added. Finally, we show that these limits are physically attainable by training a diffractive optical neural network (DONN) on the Fisher information of its photon-counting outputs, yielding measurements that saturate the NHCRB. We conclude by demonstrating image reconstructions in which the trained network estimates several amplitudes of the object simultaneously and optimally.

\textit{Imaging as multiparameter estimation.---\label{WZL_SectionII}}
We assume passive imaging with widefield illumination. Let the incoherent emission intensity of the object be $f(\mathbf{r};\boldsymbol{\theta})$. Within a square field of view (FOV) of side $L$, we parametrize this intensity using its Fourier-cosine expansion~\cite{Li2026}:
\begin{equation}
f(\mathbf{r};\boldsymbol{\theta})=a_0+\sum_{\substack{0<|\mathbf{k}|\leq k_c\\k_x, k_y \geq 0}} a_{\mathbf{k}}\cos(k_x x)\cos(k_y y),
    \label{WZL_Eq1}
\end{equation}
where $\boldsymbol{\theta}=\{a_{\mathbf{k}}\}$ is the set of $M$ unknown real amplitudes. The spatial co-ordinates and frequencies are $\mathbf{r}=(x,y)$, with $x,y\in[0,L]$, and $k_{x,y}=\pi m_{x,y}/L$, respectively, with $m_{x,y}$ denoting the non-negative integer mode indices. 
A hard numerical aperture $\text{NA}$ transmits frequencies only up to the incoherent cutoff $k_c=4\pi\,\text{NA}/\lambda$ ($\lambda$ is the wavelength)~\cite{Goodman2005}. Truncating the expansion to $\vert{}\mathbf{k}\vert{}\leq k_c$ thus captures all measurable information exactly, without discarding any observable data. This limits the number of parameters to $M_c=k_c^2L^2/4\pi$ in 2D and $M_c=k_cL/\pi$ in 1D. We also treat the background $a_0$ as known, because it carries no spatial structure and can be calibrated in advance. With this parametrization, general incoherent imaging is reduced to estimating the finite number of amplitudes $\boldsymbol{\theta}$. 

Information about these amplitudes is carried to the image plane by a quantum state of light collected by the aperture. For weak incoherent sources, at most one photon is collected within each temporal mode defined by the coherence time~\cite{Tsang2016}. Therefore,  the measurement outcome consists of independent single-photon detection events. A photon emitted from position $\mathbf{r}$ reaches the image plane in the mode $\ket{\psi_\mathbf{r}}$, the amplitude point-spread function of the aperture centered at $\mathbf{r}$. Because independent emissions are mutually incoherent, each detected photon is described by a classical mixture of these modes, weighted by the normalized source distribution:
\begin{equation}
\rho(\boldsymbol{\theta})=\frac1{a_0L^2}\int_{[0,L]^2}
f(\mathbf{r};\boldsymbol{\theta})
\ket{\psi_\mathbf{r}}\!\bra{\psi_\mathbf{r}}\ d^2\mathbf{r}.
    \label{WZL_Eq2}
\end{equation}
This state is linear in the amplitudes, meaning that $\rho(\boldsymbol{\theta})=\rho_0+\sum_{\mathbf{k}}a_{\mathbf{k}}\,\partial_{\mathbf{k}}\rho$, where $\rho_0$ is the known background state and $\partial_{\mathbf{k}}\rho\equiv\partial\rho/\partial a_{\mathbf{k}}$ are fixed, amplitude-independent operators.

\textit{Evaluation of precision bounds.---}
We evaluate the members of Eq.~\eqref{WZL_Eq3} in the low-contrast regime $a_{\mathbf{k}}\ll a_0$ of faint features on a bright background. This regime is standard for extended fluorescent objects, because positivity of $f(\mathbf{r};\boldsymbol{\theta})$ bounds the total modulation by the background, and a densely labeled scene shares it among many spatial-frequency components, leaving each amplitude individually small even when the image is not faint. Moreover, the CRBs calculated under this approximation have been found to closely match the precise CRBs even for high-contrast objects~\cite{Li2026}. 

For direct imaging (DI; position-basis measurement in the image plane), and for the quantum limit, the Fisher information matrices are diagonal and analytic (Sec.~S3):
\begin{equation}
F^{(\mathrm{DI})}_{\mathbf{k}\mathbf{k}'}=\frac{\OTF(\mathbf{k})^2}{4a_0^2}\,\delta_{\mathbf{k}\mathbf{k}'},
\qquad
Q_{\mathbf{k}\mathbf{k}'}=\frac{\OTF(\mathbf{k})}{4a_0^2}\,\delta_{\mathbf{k}\mathbf{k}'},
    \label{WZL_Eq4}
\end{equation}
with $\OTF(\mathbf{k})$ being the optical transfer function (the Fourier image of the intensity PSF), which is nearly conical with $\OTF(0)=1$ and falls monotonically to zero at the cutoff~\cite{Goodman2005}. Each amplitude thus has a per-photon variance $4a_0^2/\OTF^2$ for DI but only $4a_0^2/\OTF$ at the quantum limit (in one-dimension, the factor $4a_0^2$ in Eq.~\eqref{WZL_Eq4} becomes $2a_0^2$). Their ratio $1/\OTF(\mathbf{k})$ diverges toward the cutoff, indicating that significant precision advantage can be gained by judicious measurement design. 

Notably, our imaging model satisfies the weak commutativity condition, in which the SLDs of each amplitude in $\boldsymbol{\theta}$ commute on average with respect to $\rho(\boldsymbol{\theta})$ (Sec.~S2). Hence, the Holevo bound is identical to the QCRB~\cite{Matsumoto2002}, and is attainable with collective measurements. However, as we show next, restricting to single-copy measurements results in loss of this attainability.

The remaining member, the NHCRB, generally admits no closed form. It was first proposed by Nagaoka for two parameters \cite{Nagaoka1989} and subsequently generalized by Hayashi to $M$ parameters~\cite{Hayashi1999}. Conlon \textit{et al.} recast this bound as a semidefinite program~\cite{Conlon2021} 
\begin{align}
\cNH
&= \min_{\mathbb{L},X}
   \sum_{\mathbf{k}}
   \Tr\!\left[
      \rho(\boldsymbol{\theta})\mathbb{L}_{{\mathbf k}{\mathbf k}}
   \right]
\label{SDP1}\\[2pt]
&\text{subject to }
\begin{pmatrix}
  \mathbb{L} & X\\[-1pt]
  X^{\top}   & \mathbb{I}_K
\end{pmatrix}
\succeq 0,
\ 
\Tr\!\left[(\partial_{\mathbf k}\rho)X_{\mathbf k'}\right]
= \delta_{{\mathbf k}{\mathbf k'}}.
\nonumber
\end{align}
where $X=(X_1,\dots,X_M)^{\top}$ is a column of Hermitian observables on the photon transmitted through the aperture, $\mathbb{L}$ is an $M\times M$ matrix of Hermitian operators $\mathbb{L}_{{\mathbf k}{\mathbf k'}}$ obeying block symmetry $\mathbb{L}_{{\mathbf k}{\mathbf k'}}=\mathbb{L}^\dagger_{{\mathbf k'}{\mathbf k}}$. This problem is then a standard SDP with the block side $K(M+1)$, where $K$ is the dimension of the measurement Hilbert space.  

Although this dimension is nominally infinite, we can restrict our analysis to its subspace defined by the set of orthogonal optical modes transmitted through the aperture. This subspace is spanned by the eigenfunctions of the Gram matrix $\bra{\psi_\mathbf{r}}\psi_\mathbf{r'}\rangle$, with $\mathbf r\in[0,L]^2$, convolved with the amplitude PSF (see Sec.~S4 for a proof). We calculate the eigenspectrum of this matrix and  retain $K\sim M_c$ eigenmodes whose corresponding eigenvalues exceed a fixed small threshold. The modes below this threshold receive a vanishing share of the collected light and therefore can be neglected. The resulting SDP problem can be solved using one of the standard solvers, albeit with computational costs growing rapidly with $K$ and $M$. 

\textit{Physical architecture.---} Our optical instrument is the DONN --- a cascade of modulation surfaces, trainable pixel by pixel, which modify the phase of the incident optical signals, thereby applying a linear transformation to the input optical mode. Since their inception in 2018~\cite{Ozcan2018}, DONNs have been used for a variety of applications including deep learning, image recognition and reconstruction, and communications \cite{Chen2024}\footnote{DONNs are technically identical to multiplane light converters~\cite{Morizur2010,Fontaine2019,Zhang2023}, with the difference in terminology arising from different areas of application and, in some cases, training method.}. In our DONN, $P$ phase masks (layers) are interleaved with optical Fourier transforms, $\mathcal{F}$
, so successive masks act alternately in the spatial and pupil domains. The cascade therefore implements the unitary
\begin{equation}
V(\boldsymbol{\phi})=\mathcal{F}\,e^{i\phi_P}\cdots\,
{\mathcal{F}\,e^{i\phi_2}\,\mathcal{F}\,e^{i\phi_1}}
\label{Eq_DONN_unitary}
\end{equation}
with $\phi_i$ denoting the 2D array of phase shifts in the $i^{\text{th}}$ modulation surface. The DONN is followed by a single-photon detector array, which measures every photon in the trainable basis $\ket{v_{d}}=V^{\dagger}(\boldsymbol{\phi})\ket{d}$, where $d=(\xi,\eta)$ labels the detector pixel coordinates. A run of $N$ photons returns counts $n_{d}$ over the output pixels, with the 
expectation $\mathbb{E}[n_d]=N\,u_d(\boldsymbol{\theta})$, where
\begin{align}
u_d(\boldsymbol{\theta})
=\bra{v_d}\rho(\boldsymbol{\theta})\ket{v_d}
=u_d(0)+\sum_{\mathbf{k}\ne\mathbf{0}}
a_{\mathbf{k}}\,B_{\mathbf{k}d},
\label{Eq_DONN_probs}
\end{align}
where $B_{\mathbf{k}d}=\bra{v_d}\partial_\mathbf{k}\rho\ket{v_d}$ and $u_d(0)=a_0B_{0d}$ are independent of the parameters $\{a_{\mathbf{k}}\}$ owing to the linearity of the density matrix \eqref{WZL_Eq2} in these parameters. 
For the Poissonian photon-number distribution, the FIM is ~\cite{Tsang2016}
\begin{equation}\label{FI}
    F_{\mathbf{k}\mathbf{k}'}(\boldsymbol{\phi}) =\sum_d B_{\mathbf{k}d}B_{\mathbf{k}'d}/u_d(0),
\end{equation} 
also parameter independent. This makes the estimation precision a differentiable function of the mask phases, and enables an innovative DONN training strategy. Rather than following the two-step approach of first solving for an optimal measurement basis and then engineering optics that sort into it, we train the DONN \textit{directly} by gradient descent to minimize $\Tr[F^{-1}(\boldsymbol{\phi})]$ for the parameter set of interest. The measurement basis then emerges as a by-product.

The detector array counts are then fed into a matched linear estimator: a weighted linear regression of the observed counts on a model determined by $u_d(0)$ and the set of $\{B_{\mathbf{k}d}\}$, which can be  measured in advance with the object absent. The regression is solved by the single linear map (Sec.~S5):
\begin{equation}
\hat{\boldsymbol{\theta}}=F^{-1}B\,U_0^{-1}\!
\left(\frac{\mathbf{n}}{N}-\mathbf{u}(0)\right),
\quad U_0=\mathrm{diag}[\mathbf{u}(0)]
\label{WZL_Eq5}
\end{equation}
where $B$, $\mathbf{n}$ and $\mathbf u(0)$ are the matrix and vectors corresponding to $B_{\mathbf k d}$, $n_d$, and $u_d(0)$, respectively.
The estimator is unbiased and attains the covariance $F^{-1}/N$ at $\boldsymbol{\theta}=0$, closing the first inequality of Eq.~\eqref{WZL_Eq3}. 
\begin{figure}
    \centering
    \includegraphics[width=0.9\linewidth]{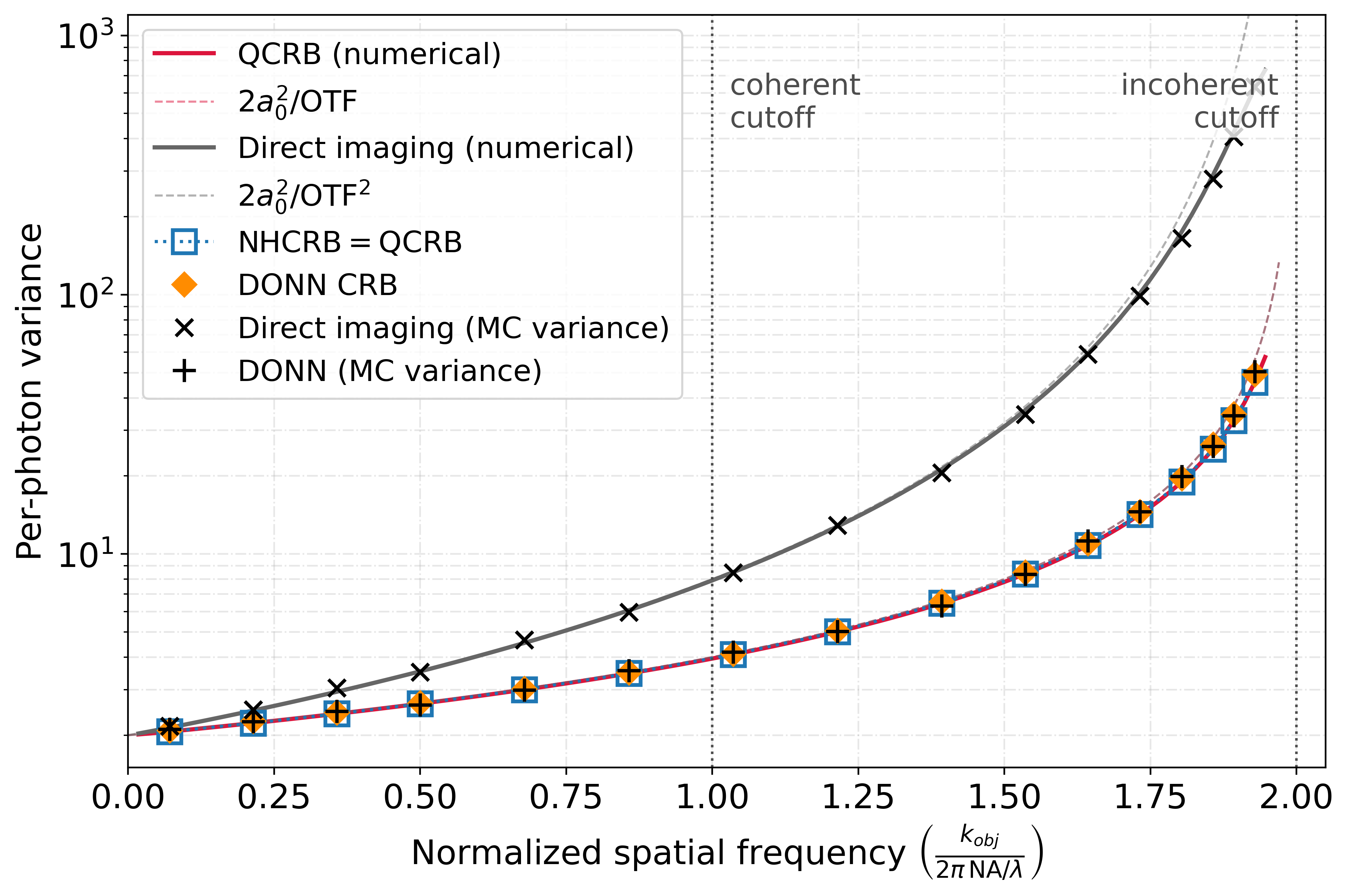}
    \caption{Variances for single-amplitude estimation and the corresponding precision bounds. The NHCRB coincides with the QCRB, and the trained 8-layer DONN saturates it at every spatial frequency. DI falls short of this benchmark. The CRBs for DI and DONN are consistent with the variance of the estimator~\eqref{WZL_Eq5} tested via Monte Carlo (MC) simulations.}
    \label{fig:WZL_Fig1.png}
\end{figure}

\begin{figure}
    \centering
    \includegraphics[width=0.9\linewidth]{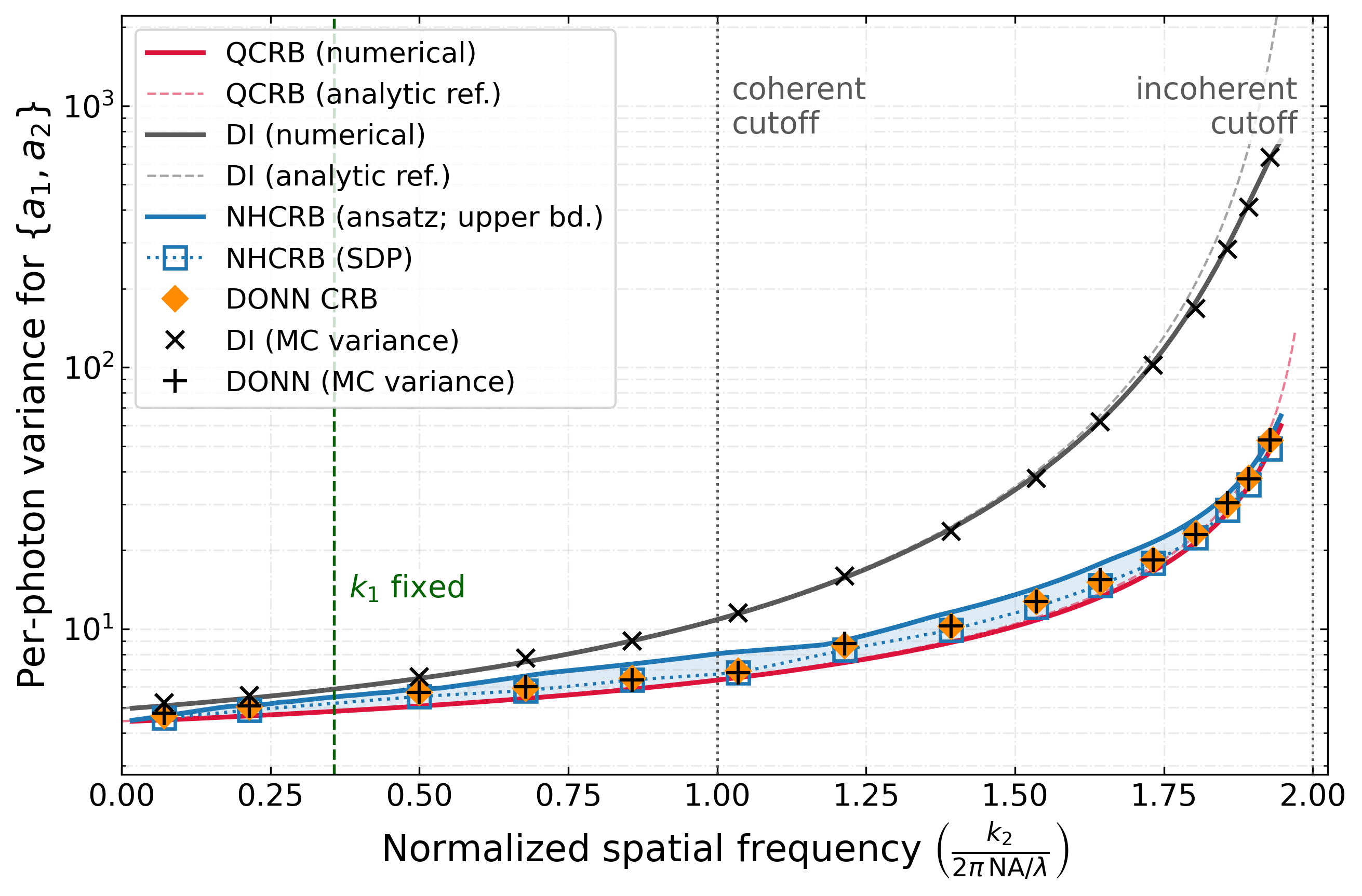}
    \caption{Two-amplitude estimation. The first spatial frequency is fixed at 0.36. The total variance of both amplitudes is plotted as a function of the second spatial frequency, with the precision bounds $\mathrm{Tr}[F^{-1}]$. The NHCRB detaches from the QCRB and reconverges near the incoherent cutoff. Shaded region shows the upper bound~\eqref{WZL_Eq6}, validating our SDP implementation, and a 10-layer DONN saturates the NHCRB across the band.}
    \label{fig:WZL_Fig2.png}
\end{figure}

\begin{figure}
    \centering
    \includegraphics[width=0.9\linewidth]{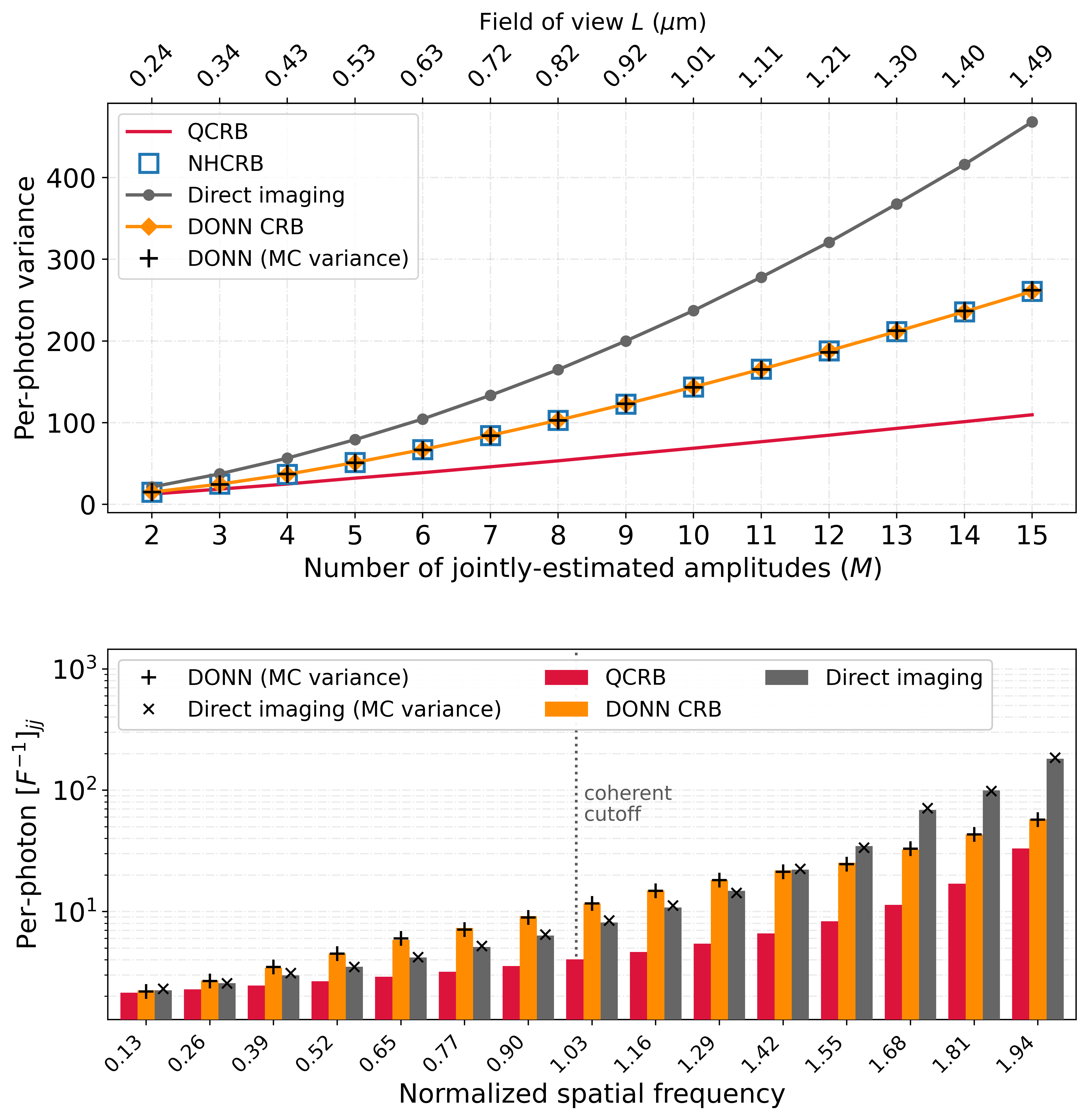}
    \caption{Total variances for multiple amplitude estimation and the corresponding precision bounds.~(a) For up to fifteen parameters equally spaced across the band, the NHCRB scales linearly, away from the QCRB. A suitably deep DONN, taken as $2M$ layers for $M$ parameters, consistently saturates the NHCRB as parameters increase.~(b) The per-parameter variances (log-y) at $M=15$ show that DONN gains advantage over DI by prioritizing higher frequency amplitudes.}
    \label{fig:WZL_Fig3.png}
\end{figure}

\textit{Results.---} Throughout our numerical studies, we consider microscopic objects emitting the wavelength of $\lambda=540$ nm, viewed through an objective with NA$=1.4$, which corresponds to $k_c=32.6\ \mu$m$^{-1}$ (minimum resolvable period 193 nm and the Rayleigh distance of 235 nm). Figures~\ref{fig:WZL_Fig1.png}--\ref{fig:WZL_Fig3.png} show the precision benchmarks in a 1D setting. Frequencies are quoted in units of $2\pi\,\mathrm{NA}/\lambda$, so the incoherent cutoff sits at $2$. The DI and DONN CRBs are each estimated in two ways: first, via the Fisher information \eqref{FI} and second, by Monte Carlo simulation, i.e.~applying the estimator \eqref{WZL_Eq5} to sampled detection counts and calculating the variance over 5000 simulation runs. Both methods show consistent results.

We begin with single-parameter estimation (Fig.~\ref{fig:WZL_Fig1.png}), by considering the object $f(x)=a_0 + a_1 \cos(k_{\text{obj}}x)$. In this case, there is no incompatibility to pay. The NHCRB returned by the SDP is identical to the QCRB at every frequency, $\cNH=\cQ$. Therefore, the hierarchy of Eq.~\eqref{WZL_Eq3} collapses onto a single attainable limit. The CRB of DI separates from the quantum benchmarks as predicted by Eq.~\eqref{WZL_Eq4}. At every swept frequency, a trained eight-layer DONN saturates the quantum limit up to small residuals arising from finite DONN depth rather than physical obstruction. We see that near the band edge, direct imaging needs many times more photons than the DONN receiver for the same precision on a single Fourier amplitude.

Incompatibility enters with the second amplitude (Fig.~\ref{fig:WZL_Fig2.png}), for $f(x)=a_0 + a_1 \cos(k_1x) + a_2 \cos(k_2x)$. We fix $k_1=0.36$ and sweep $k_2$ across the band, estimating $(a_1,a_2)$ jointly. The SLDs of distinct frequencies do not commute, and the SDP now returns $\cNH$ above $\cQ$. 
The detachment can be bounded analytically by evaluating Nagaoka's objective~\cite{Nagaoka1989, Conlon2021}, yielding (Sec.~S4), 
\begin{equation}
\cNH\ \le\ \cQ+\frac{4a_0^2\,\mathcal{W}}{\OTF(k_1)\cdot\OTF(k_2)},
\label{WZL_Eq6}
\end{equation}
with $\mathcal{W}=\min_i\{\OTF(k_i),\,1-\OTF(k_i)\}$. The right-hand side of this inequality is plotted as the upper boundary of the shaded region in Fig.~\ref{fig:WZL_Fig2.png}; we see that numerical data from the SDP are consistent with it. We also find that at every swept frequency, the trained 10-layer DONN saturates the NHCRB.

Figure~\ref{fig:WZL_Fig3.png} scales the task from two amplitudes to fifteen, jointly estimated and equally spaced across the band. We consider an object of variable size, setting $M=M_c= \lfloor k_cL/\pi \rfloor$. 
The NHCRB detaches from the QCRB monotonically in $M$. 
At every $M$, a trained DONN of $2M$ layers saturates the NHCRB. We see that all CRB benchmarks grow linearly with the number of modes when that number substantially exceeds 1 (i.e.~the object is much larger than the PSF size). This is to be expected because the intensities of distant points in the object plane are effectively independent parameters. The asymptotic ratio between the NHCRB and DI CRB in the limit of an infinitely large object, which would quantify the ultimate quantum advantage attainable for single-copy measurements, is not yet known and worth investigating in the future. 

\begin{figure*}
    \centering
    \includegraphics[width=\linewidth]{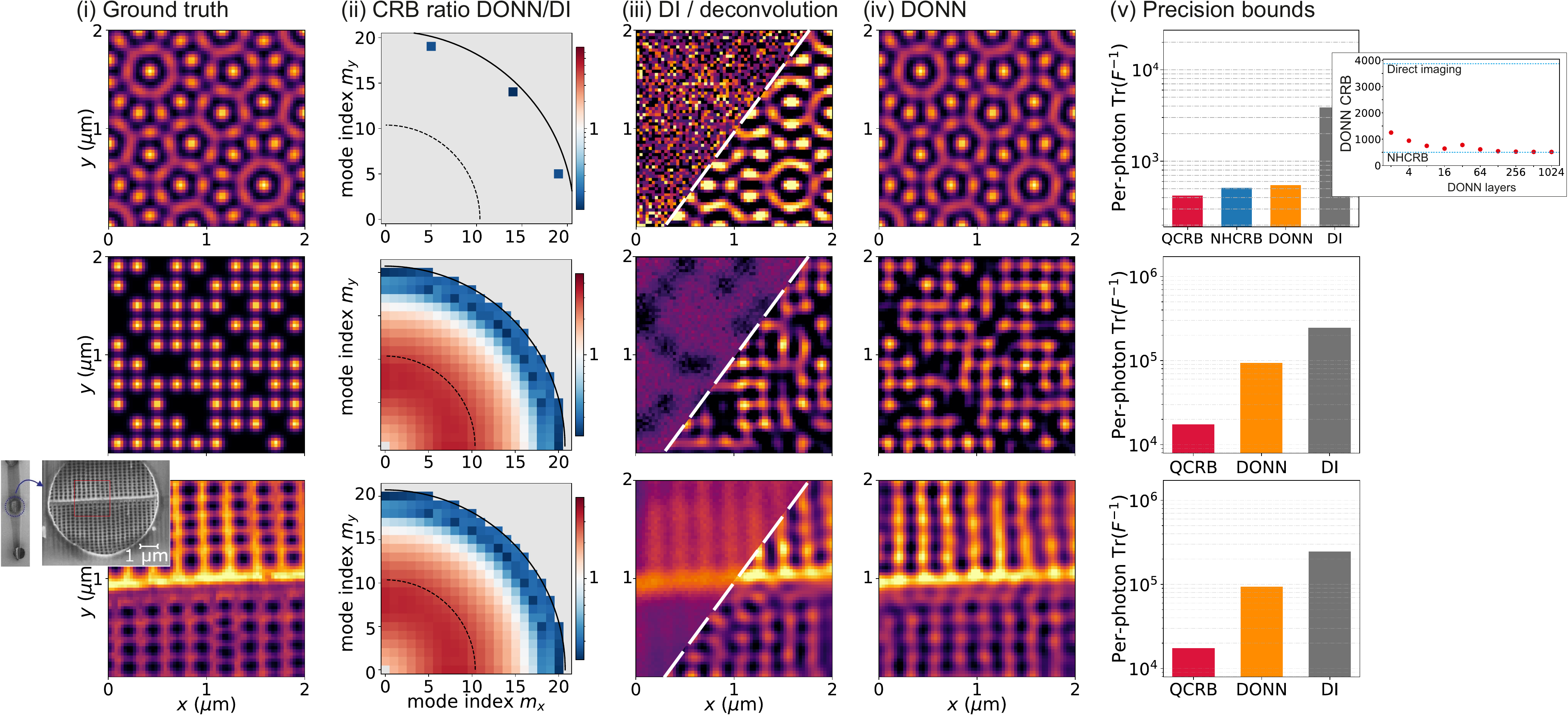}
    \caption{Image reconstructions of 2D objects with sub-Rayleigh features. Each object occupies a $2\times2\ \mu$m square field of view. Top: abstract pattern with three nonzero spatial frequency components. Middle: simulated atomic lattice. Bottom: SEM microphotograph of  \textit{Amphipleura pellucida}. 
    (i) Ground truth object. (ii) DONN/DI variance ratio for individual spatial frequency components showing DONN advantage for high frequencies.
    (iii) Direct image and its deconvolution. 
    (iv) DONN reconstruction, which recovers fine features of the object. 
    (v) Precision bars comparing per-photon $\Tr[F^{-1}]$ for the QCRB, NHCRB (computable for the first object only), DONN, and DI. Top inset: dependence of the DONN precision on the number of its layers.}
    \label{fig:WZL_Fig4.png}
\end{figure*}

\textit{Imaging arbitrary objects.---}We now generalize the framework to  objects in two dimensions, considering three $2\times2\ \mu m$ square objects in Fig.~\ref{fig:WZL_Fig4.png}. The ground truth objects are displayed in column (i).  The first object is an abstract pattern known  \textit{a priori} to contain a small set of $M=3$ nonzero spectral components near the cutoff. This small $M$ reduces the SDP block side, allowing numerical calculation of the NHCRB. The second object is a defected atomic lattice with a square pattern of 200 nm pitch. The final object is a SEM microphotograph of a frustule fragment of the diatom \textit{Amphipleura pellucida}~\cite{Sclafani2013}. This biological tissue was chosen to test the resolution of our method: the striae are about 300 nm apart and the punctae about 200 nm apart, just above the cutoff point. The mean number of photons detected was 16,384 for the first object and 40,960,000 for the second and third object. 

All DONNs had 128 layers. The DONN for the first object was trained with the prior knowledge of the three nonzero component frequencies. For the second and third objects, identical DONNs were used, trained without any prior assumptions with $M=M_c=314$ and $K=358$; these high mode counts prevented NHCRB estimation. Column (iii) displays the direct-imaging benchmark as well as a deconvolution thereof obtained by substituting the Fourier amplitudes estimated via Eq.~\eqref{WZL_Eq5} (with $\mathbf n$ being the direct image photon counts) into the Fourier expansion \eqref{WZL_Eq1}. Column (iv) shows the DONN reconstruction computed in the same way, now with $\mathbf n$ being the photon counts in the DONN outputs. Both the deconvolved DI and DONN reconstructions exhibit sub-Rayleigh features; however, the DONN yields visibly more accurate images thanks to more precise measurement of high-frequency components [column (ii)]. This advantage is further confirmed by direct comparison of the precision bounds in column (v). 
The ratios in column (ii) suggest potential benefit of hybrid schemes~\cite{Li2026}, in which low spatial frequency components are estimated by DI while those near the cutoff --- via a DONN.

Throughout the paper, we trained relatively deep DONNs to show the fundamental capability of this instrument. One may argue that such depths are difficult to realize in practice. However, as evidenced by the inset of Fig.~\ref{fig:WZL_Fig4.png} (top panel), the DONN significantly surpasses DI with as few as two layers, and reaches within a few percent of the NHCRB benchmark with as few as 8--16 layers. Such DONNs are experimentally feasible~\cite{DErrico2020, Ruan2024}. This plot also shows the attainability of the NHCRB by a DONN with sufficiently many layers.

\textit{Summary \& Discussion.---} We have recast incoherent imaging of arbitrary diffraction-limited objects as a finite multiparameter quantum estimation problem, computed the precision limit that governs single-copy measurements, and introduced a physical receiver that attains it. This is the first realizable receiver shown to reach the quantum limit on a large parameter set.  This saturation holds across the transmitted band both in one and two dimensions. Importantly, our receiver is trained without any prior knowledge of the scene, aside from a one-time, offline calibration step. Once aligned, the device operates on scenes it has not seen, akin to a camera. 

We tested our method in simulated microscopy settings. However, due to the passive nature of our method, we can expect our trained diffractive front end to be equally beneficial in astronomy, satellite imaging and remote sensing. Our method is advantageous wherever photon number is the limiting resource: fluorophores that blink and bleach, or nonrepeatable scenes with severely limited photon flux~\cite{Li2026,Lvovsky2026, Bowen2026}. Its technical realization is limited to replacing camera lenses with bespoke phase masks, which are commonly available commercially. 

Several extensions of our approach are worth exploring. First, we have assumed no prior knowledge of the sample, which makes the Fourier amplitudes a natural parametrization. When prior information is available this is no longer the case. A particularly interesting example is when the object is of limited size, in which case the optimal decomposition is set by the eigenmodes of the constrained propagation problem, and spatial frequencies beyond the cutoff can in principle be recovered through superoscillatory analytic continuation \cite{tsang2022,chang2026super}. Second, while our scheme is passive, similar optimization applies when the illumination is under control. Image scanning and structured illumination microscopy offer natural first settings, in which the illumination pattern is fixed and known; beyond these, the pattern itself can be customized via a DONN \cite{ma2026machine}, which can be trained jointly with the detection optics. An even more exciting and challenging extension is towards adaptive measurements, in which the optical network parameters are updated in real time based on prior knowledge or on preceding detection outcomes \cite{Guha2024}. Third, our work explores and optimizes the CRB, i.e.~the unweighted sum of variances of all parameters. Other figures of merit that assign unequal significance to the parameters are also worth studying: for example, we may wish to give higher weight to spatial frequency components that are close to the cutoff. Finally, we have assumed shot-noise-limited detection throughout. In many practical regimes the dominant noise is technical rather than Poissonian --- camera read noise, excess multiplication noise, dark counts --- which alters the likelihood function and hence both the Fisher information and the measurement that optimizes it. The advantages with respect to DI that can be achieved by means of trainable measurements under these conditions warrant investigation. Overall, our work shows that incorporating AI methods into quantum sensing allows attaining quantum-limited performance on previously unreachable parameter sets, giving access to a large new class of practically relevant problems. 


\textit{Acknowledgements.---} The project is funded by EPSRC Standard Grant EP/Y020596/1. AZ acknowledges a UKRI Postdoctoral Fellowship under the UK Government's Horizon Europe funding Guarantee (Q-LIMAGE, EP/Y029127/1). AW acknowledges support from Oxford's Clarendon Fund and the K. \& V. Tregidgo Scholarship. We thank Stanisław Kurdziałek, Nico Deshler, Amit Ashok, and Saikat Guha for insightful discussions. This paper was prepared in parallel with the related work of Deshler, Ashok, and Guha.

\textit{Data availability.---} The code and data required to reproduce the results and figures reported in this work are available at \href{ https://github.com/aakashwarkee/quantum-limited-DONN}{ https://github.com/aakashwarkee/quantum-limited-DONN}.

\bibliographystyle{apsrev4-2}
\bibliography{bib}

\clearpage
\onecolumngrid
\setcounter{table}{0}
\renewcommand{\thetable}{S\arabic{table}}
\renewcommand{\theequation}{S\arabic{equation}}
\setcounter{equation}{0}

\begin{center}
\textbf{\large Supplemental Material}
\end{center}
\medskip

\section*{S1. MODEL AND SIMULATION PARAMETERS}
\noindent Throughout this work, we assume a hard pupil with numerical aperture $\mathrm{NA}=1.4$ and $\lambda=540$~nm. The corresponding Rayleigh limit is $0.61\lambda/\mathrm{NA}=235$~nm. Spatial frequencies are normalized in units of $2\pi\,\mathrm{NA}/\lambda$, such that the coherent cutoff lies at $1$ and the incoherent cutoff, $k_c=4\pi\,\mathrm{NA}/\lambda$, at $2$. We set the background to $a_0=1$, and evaluate all bounds under the low-contrast assumption $a_\mathbf{k}\ll a_0$. The remaining parameters are listed in Table~S1:
\begin{table}[h]
\centering
\begin{tabular}{lccccc}
\hline\hline
 & Fig.~\ref{fig:WZL_Fig1.png} & Fig.~\ref{fig:WZL_Fig2.png} & Fig.~\ref{fig:WZL_Fig3.png} & \multicolumn{2}{c}{Fig.~\ref{fig:WZL_Fig4.png}} \\
 &  &  &  & pattern; & atoms/diatom \\
\hline
Dimension                       & 1D      & 1D      & 1D          & 2D             & 2D \\
Field of view $L$ ($\mu$m)      & $10.8$  & $10.8$  & $0.24$--$1.49$ & $2.0$ & $2.0$ \\
Field samples                   & $256$   & $256$   & $256$       & 64$\times$64   & 64$\times$64 \\
Detector samples                & $512$   & $512$   & $512$       & $16,384$             & $16,384$ \\
Mode-space dimension $K$        & $69$    & $62$    & $\approx0.8M+4$ & $128$      & $358$ \\
Parameters $M$                  & $1$     & $2$     & $2$--$15$   & $3$            & $314$ \\
DONN layers                     & $8$     & $10$    & $2M$        & $128$          & $128$ \\
SDP block side $(M{+}1)K$       & $138$   & $186$   & $\le240$    & $512$          & $112,770$ \\
Monte Carlo runs                & $5000$  & $5000$  & $5000$      & $5000$         & $5000$\\
\hline\hline
\end{tabular}
\caption{Simulation parameters for the four figures of the main text.}
\end{table}

\section*{S2. WEAK COMMUTATIVITY PROOF}

\noindent Consider the imaging model of the main text: the state $\rho(\boldsymbol{\theta})$ of Eq.~\eqref{WZL_Eq2}, with parameters $\boldsymbol{\theta}=\{a_{\mathbf{k}}\}$ and SLDs $\mathcal{L}_{\mathbf{k}}$ defined through $\partial_{\mathbf{k}}\rho=\tfrac{1}{2}(\rho\,\mathcal{L}_{\mathbf{k}}+\mathcal{L}_{\mathbf{k}}\rho)$, where $\partial_{\mathbf{k}}\rho \equiv \partial \rho/\partial a_{\mathbf{k}}$. The weak commutativity condition at $\boldsymbol{\theta}$ is
\begin{equation}
  \Tr\!\left(\rho(\boldsymbol{\theta})\,[\mathcal{L}_{\mathbf{k}},\mathcal{L}_{\mathbf{k}'}]\right)=0
  \quad\text{for all pairs } (\mathbf{k},\mathbf{k}').
  \label{WC1}
\end{equation}
Its significance is that the Holevo bound, which is the ultimate precision limit of multiparameter estimation, coincides with the QCRB if and only if Eq.~\eqref{WC1} holds. A formal proof of this is available in Ref.~\cite{Matsumoto2002}. Because the Holevo bound is asymptotically attainable by collective measurements on many copies of the state, weak commutativity promotes the QCRB from a bound to a limit that is actually reachable in this collective sense, with no tighter limit hidden between it and the NHCRB.

Here, we show that Eq.~\eqref{WC1} is satisfied for any $\boldsymbol{\theta}$. To do so, we will first prove that the SLDs are real and symmetric. Let the density operator in the position basis have the kernel  $\langle \mathbf{r}_0\vert{}\rho(\boldsymbol{\theta})\vert{}\mathbf{r}_0'\rangle$. For spatially incoherent imaging through a hard pupil, the amplitude point-spread function $\psicap(\mathbf{r})$ is real. Thus, for an object distribution $f(\mathbf{r};\boldsymbol{\theta})$, we can write
\begin{equation}
\langle \mathbf{r}_0\vert{}\rho(\boldsymbol{\theta})\vert{}\mathbf{r}_0'\rangle = \int d^2\mathbf{r} \, f(\mathbf{r};\boldsymbol{\theta}) \, \psicap(\mathbf{r}_0-\mathbf{r})\psicap(\mathbf{r}_0'-\mathbf{r}).
\end{equation}
Since $f(\mathbf{r};\boldsymbol{\theta})$ and $\psicap(\mathbf{r})$ are real, so is $\langle \mathbf{r}_0\vert{}\rho(\boldsymbol{\theta})\vert{}\mathbf{r}_0'\rangle$. Furthermore, because the product is invariant under the exchange $\mathbf{r}_0 \leftrightarrow \mathbf{r}_0'$, we have $\langle \mathbf{r}_0\vert{}\rho(\boldsymbol{\theta})\vert{}\mathbf{r}_0'\rangle = \langle \mathbf{r}_0'\vert{}\rho(\boldsymbol{\theta})\vert{}\mathbf{r}_0\rangle$. Hence, $\rho(\boldsymbol{\theta})$ is a real symmetric matrix in the position basis. Differentiating with respect to amplitude $a_{\mathbf{k}}$ gives
\begin{equation}
    \langle \mathbf{r}_0\vert{}\partial_{\mathbf k}\rho\vert{}\mathbf{r}_0'\rangle = \int d^2\mathbf{r} \, \partial_\mathbf{k} f(\mathbf{r};\boldsymbol{\theta}) \, \psicap(\mathbf{r}_0-\mathbf{r})\psicap(\mathbf{r}_0'-\mathbf{r}),
\end{equation}
which preserves these properties. We thus have $(\partial_{\mathbf k}\rho)^{\top}=\partial_{\mathbf k}\rho$ and $(\partial_{\mathbf k}\rho)^*=\partial_{\mathbf k}\rho$. 

Now, we diagonalize $\rho(\boldsymbol{\theta})$. The spectral theorem guarantees a real orthonormal eigenbasis: $\rho(\boldsymbol{\theta}) = O\Lambda O^{\top}$, where $\Lambda=\operatorname{diag}(\lambda_1,\lambda_2,\ldots)$ and $O$ is a real orthogonal matrix. In this basis, the derivative operator $\widetilde{\partial_{\mathbf k}\rho} = O^{\top}(\partial_{\mathbf k}\rho)O$ remains real and symmetric. For $(m,n)$ such that $\lambda_m+\lambda_n>0$, the defining equation for the SLD, written in this real eigenbasis, yields \cite{Helstrom1969}
\begin{equation}
    (\widetilde{\mathcal{L}}_{\mathbf{k}})_{mn} = \frac{2(\widetilde{\partial_{\mathbf k}\rho})_{mn}}{\lambda_m+\lambda_n};
\end{equation}
both the numerator and denominator in the right-hand side are real and symmetric, so $\widetilde{\mathcal{L}}_{\mathbf{k}}$ is likewise real symmetric. For indices satisfying $\lambda_m=\lambda_n=0$, the conventional minimum-norm choice is $(\widetilde{\mathcal{L}}_{\mathbf{k}})_{mn}=0$, which is also real symmetric. This property is preserved upon transforming back to the original basis, $\mathcal{L}_{\mathbf{k}} = O\widetilde{\mathcal{L}}_{\mathbf{k}} O^{\top}$.

Because $\rho(\boldsymbol{\theta})$ and its SLDs are all real symmetric, the trace of their product is invariant under transposition:
\begin{equation}
\Tr\!\left(\rho(\boldsymbol{\theta})\,\mathcal{L}_{\mathbf{k}}\mathcal{L}_{\mathbf{k'}}\right) =\Tr\!\left((\rho(\boldsymbol{\theta})\,\mathcal{L}_{\mathbf{k}}\mathcal{L}_{\mathbf{k'}})^{\top}\right) =\Tr\!\left(\mathcal{L}_{\mathbf{k'}}\mathcal{L}_{\mathbf{k}}\rho(\boldsymbol{\theta})\right) =\Tr\!\left(\rho(\boldsymbol{\theta})\,\mathcal{L}_{\mathbf{k'}}\mathcal{L}_{\mathbf{k}}\right),
\label{WC2}
\end{equation}
so the commutator trace vanishes identically, and Eq.~\eqref{WC1} holds without any assumption on the contrast. Remarkably, any fixed pupil phase variation, such as an aberration or defocusing, changes $\rho(\boldsymbol{\theta})$ only by a
$\boldsymbol{\theta}$-independent unitary and leaves this property intact.


\section*{S3. CRAM\'ER-RAO BOUNDS IN THE LOW-CONTRAST APPROXIMATION}\label{app:qfi}
\noindent In this section, we derive expressions for the QFI and for the FI of direct imaging quoted in Eq.~\eqref{WZL_Eq4} of the main text, following the route of Ref.~\cite{Li2026}. Throughout, $c_{\mathbf{k}}(\mathbf{r})\equiv\cos(k_xx)\cos(k_yy)$ denotes the cosine component of Eq.~\eqref{WZL_Eq1} of the main text at wavevector $\mathbf{k}$, and $\rho_0$ and $\partial_{\mathbf{k}}\rho$ are the background state and derivative operators defined as:
\begin{gather}
\rho_0 = \frac{1}{L^2}
\int_{[0,L]^2} d^2\mathbf{r}\,
\ket{\psi_\mathbf{r}}\!\bra{\psi_\mathbf{r}}, \label{S00}\\
\partial_{\mathbf{k}}\rho = \frac{1}{a_0L^2}
\int_{[0,L]^2} d^2\mathbf{r}\,
\cos(k_x x)\cos(k_y y)\,
\ket{\psi_\mathbf{r}}\!\bra{\psi_\mathbf{r}}.\label{S01}
\end{gather}
The image-plane mode of a source at $\mathbf{r}$ is
$\ket{\psi_\mathbf{r}}=\int d^2\mathbf{r}_0\, \psicap(\mathbf{r}_0 -\mathbf{r})\ket{\mathbf{r}_0}$,
where $\psicap(\mathbf{r})$ is the normalized amplitude point-spread function (APSF). In two dimensions, for the hard circular pupil, $\psicap(\mathbf{r})=J_1(\kc r/2)/(\sqrt{\pi}\,r)$, where $r=\abs{\mathbf{r}}$ and $J_1$ is the Bessel function of the first kind; its Fourier transform is flat over the pupil, $\tilde{\psicap}(\mathbf{k})=4\sqrt{\pi}/\kc$ for $\abs{\mathbf{k}}\le\kc/2$, and zero otherwise~\cite{Goodman2005}. The intensity PSF is $p(\mathbf{r})=|\psicap(\mathbf{r})|^2$ with $\int p(\mathbf{r})\,d^2\mathbf{r}=1$, and its Fourier transform is the optical transfer function $\OTF(\mathbf{k})$, with $\OTF(\mathbf{0})=1$. Because $\tilde{\psicap}$ is a top-hat, its square is proportional to itself, and the self-convolution of the APSF reproduces the APSF: 
\begin{equation}
\int d^2\mathbf{r}_0\,\psicap(\mathbf{r}-\mathbf{r}_0)
\psicap(\mathbf{r}'-\mathbf{r}_0)
=\frac{4\sqrt{\pi}}{k_c}\,\psicap(\mathbf{r}-\mathbf{r}'),
\label{S1}
\end{equation}
which also gives the mode overlaps
$\bra{\psi_\mathbf{r}}{\psi_{\mathbf{r}'}}\rangle
=(4\sqrt{\pi}/k_c)\psicap(\mathbf{r}-\mathbf{r}')$. 
Furthermore, let $\mathcal{C}$ denote the convolution operator with the intensity PSF,
$(\mathcal{C}g)(\mathbf{r})\equiv\int d^2\mathbf{r}'\,p(\mathbf{r}-\mathbf{r}')\,g(\mathbf{r}')$.
The cosine components are its eigenfunctions: $c_{\mathbf{k}}$ is a superposition of four plane waves with wavevectors $(\pm k_x,\pm k_y)$, all sharing the modulus $\abs{\mathbf{k}}$, and thus
\begin{equation}
(\mathcal{C}c_{\mathbf{k}})(\mathbf{r})
=\int d^2\mathbf{r}'\,c_{\mathbf{k}}(\mathbf{r}')\,p(\mathbf{r}-\mathbf{r}')
=\OTF(\mathbf{k})\,c_{\mathbf{k}}(\mathbf{r}),
\label{S2}
\end{equation}
up to corrections from the edges of the field of view, which are quantified at the end of this section.\\\\
\textit{Quantum Fisher information.}---The QFI matrix is given by \cite{Helstrom1969}:
\begin{equation}
Q_{\mathbf{k}\mathbf{k'}}=\mathrm{Re}\,\Tr\!\bigl[\rho\,\mathcal{L}_{\mathbf{k}}\mathcal{L}_{\mathbf{k}'}\bigr],
\label{S3}
\end{equation}
where the symmetric logarithmic derivative (SLD) $\mathcal{L}_{\mathbf{k}}$ associated with the
amplitude $a_{\mathbf{k}}$ is defined through
$\partial_{\mathbf k}\rho=\tfrac{1}{2}\bigl(\rho\,\mathcal{L}_{\mathbf{k}}+\mathcal{L}_{\mathbf{k}}\rho\bigr)$. For the finite-FOV state at arbitrary contrast, the SLDs can be obtained from the spectral decomposition of $\rho(\boldsymbol\theta)$, but no closed-form expression is known. To obtain analytic results, we evaluate the SLD equation under the low-contrast approximation: the intensity is taken spatially uniform, and $\rho(\boldsymbol{\theta})$ is replaced by the background state $\rho_0$. We now show that the SLD is proportional to the derivative operator,
\begin{equation}
\mathcal{L}^{(\mathrm{appr})}_{\mathbf{k}}
=\frac{k_c^2L^2}{16\pi}\,\partial_{\mathbf{k}}\rho.
\label{S4}
\end{equation}
We verify Eq.~\eqref{S4} directly by substitution. Using Eq.~\eqref{S1} twice,
\begin{align}
\rho_0\,\mathcal{L}^{(\mathrm{appr})}_{\mathbf{k}}
&=\frac{k_c^2}{16\pi a_0L^2}\int\! d^2\mathbf{r}'\,
c_{\mathbf{k}}(\mathbf{r}')
\Bigl[\int\! d^2\mathbf{r}\,
\bra{\psi_\mathbf{r}}{\psi_{\mathbf{r}'}}\rangle\,
\ket{\psi_\mathbf{r}}\Bigr]\bra{\psi_{\mathbf{r}'}}
\nonumber\\
&=\frac{k_c^2}{16\pi a_0L^2}\,\frac{16\pi}{k_c^2}
\int d^2\mathbf{r}'\,c_{\mathbf{k}}(\mathbf{r}')
\ket{\psi_{\mathbf{r}'}}\!\bra{\psi_{\mathbf{r}'}}
=\partial_{\mathbf{k}}\rho,
\label{S5}
\end{align}
where the bracketed integral was evaluated in the position basis, $\int d^2\mathbf{r}\,
\psicap(\mathbf{r}_0-\mathbf{r})\psicap(\mathbf{r}-\mathbf{r}')
=(4\sqrt{\pi}/k_c)\psicap(\mathbf{r}_0-\mathbf{r}')$, such that each of the two APSF factors collapse by one use of Eq.~\eqref{S1}. Similarly, $\mathcal{L}^{(\mathrm{appr})}_{\mathbf{k}}\rho_0
=\partial_{\mathbf{k}}\rho$. Hence, Eq.~\eqref{S4} solves the SLD equation under the low-contrast assumption.\\\\
With the SLD in hand, the QFI in Eq.~\eqref{S3} reduces to a double integral:
\begin{align}
Q_{\mathbf{k}\mathbf{k}'}&\approx\mathrm{Re}\,\Tr\left[\partial_{\mathbf k}\rho\,\mathcal{L}^{(\mathrm{appr})}_{\mathbf{k}'}\right]\\
&=\frac{k_c^2L^2}{16\pi}\,
\Tr\!\bigl[\partial_{\mathbf k}\rho\,\partial_{\mathbf{k}'}\rho\bigr]\\
&=\frac{k_c^2L^2}{16\pi}\frac{1}{(a_0L^2)^2}
\!\int\!\! d^2\mathbf{r}\!\int\!\! d^2\mathbf{r}'\,
c_{\mathbf{k}}(\mathbf{r})\,c_{\mathbf{k'}}(\mathbf{r}')\,
|{\bra{\psi_\mathbf{r}}{\psi_{\mathbf{r}'}}\rangle}|^2.
\label{S6}
\end{align}
By Eq.~\eqref{S1} the squared overlap is $(16\pi/k_c^2)\,p(\mathbf{r}-\mathbf{r}')$, so the prefactor of Eq.~\eqref{S4} cancels out. The $\mathbf{r}'$ integral is then the blurring identity Eq.~\eqref{S2}, and the $\mathbf{r}$ integral is the orthogonality of the cosine components on the field-of-view, $\int c_{\mathbf{k}}c_{\mathbf{k}'}\,d^2\mathbf{r}=(L^2/4)\,\delta_{\mathbf{k}\mathbf{k}'}$ for $k_x,k_y>0$. Hence
\begin{equation}
Q_{\mathbf{k}\mathbf{k}'}=\frac{\OTF(\mathbf{k})}{4a_0^2}\,\delta_{\mathbf{k}\mathbf{k}'},
\label{S7}
\end{equation}
the second expression of Eq.~\eqref{WZL_Eq4} of the main text. If exactly one Cartesian component vanishes, then the corresponding Cramér–Rao bound is reduced by a factor of two. The one-dimensional model of the frequency sweeps follows similarly, with sinc APSF and $\mathcal{L}^{(\mathrm{appr})}_{k}=(\kc L/2\pi)\,\partial_{k}\rho$, giving
$Q_{kk'}=\OTF(k)/(2a_0^2)\,\delta_{kk'}$.\\\\
\textit{Fisher information of direct imaging.}---Direct imaging measures the position of each photon via the position-basis POVM $\{\ket{\mathbf{r}}\!\bra{\mathbf{r}}\,
d^2\mathbf{r}\}$. The probability density of detecting a photon at $\mathbf{r}$ is therefore the diagonal matrix element of the state:
\begin{equation}
u(\mathbf{r};\boldsymbol{\theta})
=\bra{\mathbf{r}}\rho(\boldsymbol{\theta})\ket{\mathbf{r}}
=\bra{\mathbf{r}}\rho_0\ket{\mathbf{r}}
+\sum_{\mathbf{k}}a_{\mathbf{k}}
\bra{\mathbf{r}}\partial_{\mathbf{k}}\rho\ket{\mathbf{r}}.
\label{S8}
\end{equation}
The background term is flat, $\bra{\mathbf{r}}\rho_0\ket{\mathbf{r}} =L^{-2}\!\int d^2\mathbf{r}'\,p(\mathbf{r}-\mathbf{r}')=1/L^2$, because the intensity PSF is normalized. The derivative terms follow from the blurring identity Eq.~\eqref{S2},
$\bra{\mathbf{r}}\partial_{\mathbf{k}}\rho\ket{\mathbf{r}}
=\OTF(\mathbf{k})\,c_{\mathbf{k}}(\mathbf{r})/(a_0L^2)$, such that
\begin{equation}
u(\mathbf{r};\boldsymbol{\theta})
=\frac{1}{L^2}\Bigl[1+\sum_{\mathbf{k}}\OTF(\mathbf{k})\,
\frac{a_{\mathbf{k}}}{a_0}\,c_{\mathbf{k}}(\mathbf{r})\Bigr],
\label{S9}
\end{equation}
every cosine component is attenuated by its transfer function value, while the background is left untouched. For shot-noise-limited detection, the FI per photon is \cite{Tsang2016}:
\begin{equation}
F^{(\mathrm{DI})}_{\mathbf{k}\mathbf{k}'}=\int d^2\mathbf{r}\,\frac1u\,\frac{\partial u}{\partial a_{\mathbf{k}}}\frac{\partial u}{\partial a_{\mathbf{k}'}}
\label{S10}
\end{equation}
By Eq.~\eqref{S8} the derivatives are exact and parameter independent, $\partial u/\partial a_{\mathbf{k}}
=\bra{\mathbf{r}}\partial_{\mathbf{k}}\rho\ket{\mathbf{r}}$, while the denominator is evaluated with the same low contrast approximation as above, $u\approx\bra{\mathbf{r}}\rho_0\ket{\mathbf{r}}=1/L^2$. The
integral then reduces to the orthogonality of the cosine components:
\begin{equation}
F^{(\mathrm{DI})}_{\mathbf{k}\mathbf{k}'}=\frac{\OTF(\mathbf{k})^2}{4a_0^2}\,\delta_{\mathbf{k}\mathbf{k}'},
\label{S11}
\end{equation}
which reproduces the first equality of Eq.~\eqref{WZL_Eq4} of the main text. The optical transfer
function enters the FI squared, but the QFI \eqref{S7} only linearly, for the following reason.
In direct imaging, the signal carried by each cosine component of the detected intensity is
attenuated by $\OTF(\mathbf{k})$ [Eq.~\eqref{S9}], while the shot noise against which it must
be resolved is set by the flat background, which the aperture transmits in full. Since the
Fisher information \eqref{S10} is quadratic in the signal and inversely proportional to the noise, each
amplitude costs a per-photon variance $4a_0^2/\OTF^2$. In the quantum bound, by contrast, the
blur enters only once, through the mode overlap
$\abs{\braket{\psi_{\mathbf{r}}|\psi_{\mathbf{r}'}}}^2$ in Eq.~\eqref{S6}, giving $4a_0^2/\OTF$. We note that Eqs.~\eqref{S8} and \eqref{S10} apply unchanged to any receiver that counts photons in a discrete set of output modes: the position kets are simply replaced by the output modes of the device, a structure we exploit when constructing the measurement of the main text.\\\\
We also emphasize that these results rely on two approximations of different origin:
\begin{enumerate}[label=(\roman*)]
    \item Low-contrast replacements $\rho\to\rho_0$ in Eq.~\eqref{S3} and $u\to\bra{\mathbf{r}}\rho_0\ket{\mathbf{r}}$ in  Eq.~\eqref{S10}, accurate to first order in $a_{\mathbf{k}}/a_0$.
    \item Neglecting the finite extent of the field of view in Eqs.~\eqref{S1} and~\eqref{S2}. These identities are exact if the object is evenly extended beyond the FOV (equivalently, for an unbounded FOV), in which case the transmitted field occupies exactly $K$ discrete plane-wave modes and the cosine components are exact eigenfunctions of $\mathcal{C}$. The hard edges of the finite FOV mix these modes: the identities miscount the modes engaged at frequency $\mathbf{k}$ by $O(1)$ out of ${\sim}K\,\OTF(\mathbf{k})$, giving a relative error of $[K\,\OTF(\mathbf{k})]^{-1}\sim\lambda/(\mathrm{NA}\,L\,\OTF(\mathbf{k}))$. This is a few percent in mid-band, but order unity for components within ${\sim}\lambda/(\mathrm{NA}\,L)$ of the cutoff.
\end{enumerate}
All bounds shown in the figures are evaluated numerically on the discretized pupil under the low-contrast assumption alone~\cite{Li2026}, without invoking approximation (ii); the analytic forms are shown for reference, and their deviation from the numerical curves near the cutoff is this finite-FOV edge effect.

\section*{S4. NAGAOKA-HAYASHI SEMIDEFINITE PROGRAM}\label{app:sdp}
\noindent In this section, we show our implementation of the NHCRB in Figs.~\ref{fig:WZL_Fig1.png}--\ref{fig:WZL_Fig4.png}, and derive the analytic upper bound quoted in Eq.~\eqref{WZL_Eq6} of the main text.\\\\
\medskip\noindent\textit{The semidefinite program (SDP).}---
We first prove the statement made in the main text, that the dimension of the detection Hilbert space can be reduced to the subset spanned by
\begin{equation}
\ket{\omega_n}=\frac1{\sqrt{\lambda_n}}\int g_n(\mathbf r)\ket{\psi_{\mathbf r}}d{\mathbf r},
\end{equation}
which is obtained from the eigenvalues $\lambda_n$ and eigenfunctions $g_n(\mathbf{r'})$ of the Gram matrix  $G(\mathbf r,\mathbf r')=\bra{\psi_\mathbf{r}}\psi_\mathbf{r'}\rangle$ convolved with the amplitude PSF. We need to show that this set  is orthonormal and sufficient to represent any state of the form \eqref{WZL_Eq2}. For orthonormality, we first notice that the eigenvalues are real because $G(\mathbf r,\mathbf r')$ is Hermitian. Hence
\begin{equation}\label{GramON}
    \braket{\omega_n|\omega_m}=\frac1{\sqrt{\lambda_n\lambda_m}}\iint g_n(\mathbf r)g_m(\mathbf r')\braket{\psi_{\mathbf r}|\psi_{\mathbf r'}}d{\mathbf r}d{\mathbf r'}=\frac1{\sqrt{\lambda_n\lambda_m}}\iint g^*_n(\mathbf r)g_m(\mathbf r')G(\mathbf r,\mathbf r')d{\mathbf r}d{\mathbf r'}.
\end{equation}
Now, because $g_m(\mathbf r')$ is an eigenfunction of $G(\mathbf r,\mathbf r')$, we have $\int g_m(\mathbf r')G(\mathbf r,\mathbf r')d{\mathbf r'}=\lambda_m g_m(\mathbf r)$ and because $\{g_i(\mathbf r)\}$ are orthonormal, the double integral in Eq.~\eqref{GramON} becomes $\delta_{nm}$. We further observe that 
\begin{equation}\sum_n\sqrt{\lambda_n}g^*_n(\mathbf r)\ket{\omega_n}=\sum_ng^*_n(\mathbf r)\int g_n(\mathbf r')\ket{\psi_{\mathbf r'}}d{\mathbf r'}=\int\left[\sum_ng^*_n(\mathbf r) g_n(\mathbf r')\right]\ket{\psi_{\mathbf r'}}d{\mathbf r'}=\ket{\psi_{\mathbf r}}\end{equation}
because the sum in the square brackets is $\delta(\mathbf r-\mathbf r')$ due to the resolution of the identity. In other words, the PSF $\ket{\psi_{\mathbf r}}$ for any ${\mathbf r}$ can be written as a linear combination of $\{\ket{\omega_n}\}$, which means that mixtures \eqref{WZL_Eq2} of PSFs can also be expressed in this basis.\\\\
To implement the SDP, we make the following reductions to keep the program computable:
\begin{enumerate}[label=(\roman*)]
\item \textit{Support projection.} We selected the eigenfunctions $g_n(\mathbf r)$ with the eigenvalues exceeding a threshold, which was set at $10^{-12}$ in all cases except the first object in Fig.~4. This resulted in the number $K$ of eigenfunctions with $K\sim M_c$; both numbers correspond to the space- bandwidth product of the viewfield and the aperture. For the first object in Fig.~4, the threshold was increased to $10^{-4}$, resulting in approximately halving the value of $K$; this was necessary to accommodate the problem to the available computational resources. 
\item \textit{Real representation.} As shown in Sec.~S2, $\rho(\boldsymbol{\theta})$ and all
$\partial_{\mathbf{k}}\rho$ are real symmetric. The minimizer can then be taken real as well. All $X_{\mathbf{k}}$ and
$\mathbb{L}_{\mathbf{k}\mathbf{k}'}$ are therefore restricted to real symmetric matrices, roughly halving the number of scalar variables.
\item \textit{Rescaling of the derivative operators.} The Frobenius norms of the derivative
operators fall off toward the band edge, $\Tr[(\partial_{\mathbf{k}}\rho)^2]\propto
\OTF(\mathbf{k})$, so a joint program over low- and high-frequency amplitudes spans several
orders of magnitude in scale. We therefore substitute
$\partial_{\mathbf{k}}\rho\to\partial_{\mathbf{k}}\rho/\lVert\partial_{\mathbf{k}}\rho\rVert$,
which rescales the amplitudes as
$a_{\mathbf{k}}\to\lVert\partial_{\mathbf{k}}\rho\rVert\,a_{\mathbf{k}}$, and compensate in the
objective with the weights $w_{\mathbf{k}}=\lVert\partial_{\mathbf{k}}\rho\rVert^{-2}$:
\begin{equation}
\cNH=\min_{\mathbb{L},\,X}\ \sum_{\mathbf{k}} w_{\mathbf{k}}
\Tr\!\bigl[\rho(\boldsymbol{\theta})\,\mathbb{L}_{\mathbf{k}\mathbf{k}}\bigr].
\label{SDP2}
\end{equation}
This is an exact reparametrization, and we found it essential: without it, the first-order
solver stalls on the wide dynamic range and can report spurious values.
\end{enumerate}
We solved the one-dimensional programs (block sides up to $240$, Figs.~1--3) with the interior-point solver CLARABEL, and
cross-checked against the first-order solver SCS (up to $6\times10^4$ iterations at tolerance $10^{-9}$); the two solvers agree to at least four significant digits. For the two-dimensional program of Fig.~4 (block side $512$), whose memory requirement is beyond the practical reach of the interior-point solver, we used SCS at tolerance $10^{-4}$ with up to $2\times10^4$ iterations. We can verify the consistency of our SDP implementation using three physically meaningful results. First, for a single parameter the NHCRB must equal the QCRB, and the SDP reproduces this equality at every frequency of Fig.~\ref{fig:WZL_Fig1.png} to solver precision. Second, for two parameters the returned value must lie between the QCRB and the SLD-ansatz upper bound derived below; this sandwich holds at every swept point of Fig.~\ref{fig:WZL_Fig2.png}. Third, the bound must grow monotonically as parameters are  
added, as it does in Fig.~\ref{fig:WZL_Fig3.png}.

\medskip\noindent\textit{SLD-ansatz upper bound for two parameters.---} Nagaoka's original bound for two parameters is given by~\cite{Nagaoka1989,Conlon2021}:
\begin{align} 
\cNH = \min_{X_1,X_2} \Bigl\{ &\Tr\bigl[\rho\,(X_1^2+X_2^2)\bigr] + \TrAbs\,\rho\,[X_1,X_2] \Bigr\}, \label{NAG1} \\ \text{subject to } &\Tr[(\partial_{k_j}\rho)\,X_{j'}] = \delta_{jj'} \quad \text{for } j,j'\in\{1,2\}\label{Nagaoka_constraint},
\end{align}
where $j,j'\in\{1,2\}$ label the two estimated amplitudes at the distinct frequencies $k_1\neq k_2$, both transmitted by the aperture, $0<k_j\le\kc$, and $\TrAbs[A]$ is the sum of the moduli of the eigenvalues of $A$. This expression is also a direct reduction of the SDP given in main text's Eq.~\eqref{SDP1} for two parameters~\cite{Conlon2021}. The SDP output must lie between the QCRB and the derived upper bound in this section. Consider the following ansatz for $X_1$ and $X_2$,
\begin{equation}
X_j=\sum_{j'}\,[Q^{-1}]_{j'j}\,\mathcal{L}_{j'}.
\label{SDP3}
\end{equation}
Substituting Eq.~\eqref{SDP3} in Eq.~\eqref{Nagaoka_constraint}, we see that this ansatz satisfies the unbiasedness constraint via $\Tr[(\partial_{k_j}\rho)\,\mathcal{L}_{j'}]=Q_{jj'}$, by the definition of the quantum FIM. Simplifying Eq.~\eqref{NAG1} using this ansatz thereby shows that the first term evaluates to $\Tr[Q^{-1}]=\cQ$. The excess over the QCRB is therefore due to the second term representing the single-copy penalty. To evaluate it, we start by simplifying the commutator of $X_1$ and $X_2$ as $[X_1,X_2]=\det(Q^{-1})\,[\mathcal{L}_1,\mathcal{L}_2]$. Substituting this commutator in Eq.~\eqref{NAG1} and using the low-contrast assumption, $\rho\to\rho_0$, gives
\begin{equation}
\cNH\ \le\ \cQ+\frac{\TrAbs\bigl(\rho_0\,[\mathcal{L}_1,\mathcal{L}_2]\bigr)}{\det Q}.
\label{SDP4}
\end{equation}
Reducing Eq.~\eqref{S7} to one-dimension by replacing $4a_0^2$ with $2a_0^2$ and calculating $\det Q$ for the two amplitudes, we get:
\begin{equation}
\cNH\ \le\ \cQ+\frac{4a_0^4}{\text{OTF}(k_1)\cdot\text{OTF}(k_2)}\cdot\TrAbs\bigl(\rho_0\,[\mathcal{L}_1,\mathcal{L}_2]\bigr).
\label{SDP5}
\end{equation}
To obtain the closed-form expression quoted in Eq.~\eqref{WZL_Eq6}, we must now evaluate $\TrAbs\bigl(\rho_0\,[\mathcal{L}_1,\mathcal{L}_2]\bigr)$. For a normalized amplitude PSF in 1D, the pupil modes $\ket{k}$ transmitted by the aperture, $\abs{k}\le\kc/2$, obey $\braket{k|\psi_x}=\sqrt{2\pi/\kc L}\,e^{-ikx}$. Substituting this into the one-dimensional versions of Eqs.~\eqref{S00} and \eqref{S01} gives
\begin{align}
\braket{k|\rho_0|k'}&=\frac{2\pi}{\kc L}\,\delta_{kk'},\label{SDP_rhoID}\\
\qquad
\braket{k|\partial_{k_j}\rho|k'}
&=\frac{\pi}{\kc L a_0}\bigl(\delta_{k-k',\,k_j}+\delta_{k-k',\,-k_j}\bigr).
\label{SDPmel}
\end{align} 
Equation~\eqref{SDP_rhoID} reduces $\TrAbs\bigl(\rho_0\,[\mathcal{L}_1,\mathcal{L}_2]\bigr)$ to $\left(\frac{2\pi}{\kc L}\right)\TrAbs\bigl(\,[\mathcal{L}_1,\mathcal{L}_2]\bigr)$. We are now left with evaluating only the term $\TrAbs\bigl(\,[\mathcal{L}_1,\mathcal{L}_2]\bigr)$. To do this, we again use the SLD equation $\partial_{k_j}\rho=\tfrac12(\rho_0\mathcal{L}_j+\mathcal{L}_j\rho_0)$ and write $\partial_{k_j}\rho=(2\pi/\kc L)\,\mathcal{L}_j$. This gives Eq.~\eqref{SDPmel} in terms of the SLDs $\mathcal{L}_j$, which can be written in the operator form,
\begin{equation}
\mathcal{L}_j=\frac{S_j+S_j^{\dagger}}{2a_0},
\qquad
S_j=\!\!\!\sum_{\;\abs{k},\,\abs{k+k_j}\le\kc/2}\!\!\!\ket{k+k_j}\!\bra{k}.
\label{SDPsld}
\end{equation}
Here, $S_j$ acts on $\ket{k}$ as
\begin{equation}
S_j\ket{k}=
\begin{cases}
\ket{k+k_j}, & \abs{k+k_j}\le \kc/2,\\[2pt]
0, & \text{otherwise}.
\end{cases}
\label{SDPaction}
\end{equation}
Substituting Eq.~\eqref{SDPsld} into $[\mathcal{L}_1,\mathcal{L}_2]$ produces four terms, of which the like-signed pairs vanish. This is because $S_1S_2$ and $S_2S_1$ both take $\ket{k}$ to $\ket{k+k_1+k_2}$, and by Eq.~\eqref{SDPaction} either ordering is blocked only if $\abs{k+k_2}>\kc/2$ or $\abs{k+k_1}>\kc/2$, respectively; for $k_1,k_2>0$ both conditions are met exactly when $\abs{k+k_1+k_2}>\kc/2$. The two orderings are therefore blocked together, giving $[S_1,S_2]=0$ and likewise $[S_1^{\dagger},S_2^{\dagger}]=0$. Simplifying, we get:
\begin{equation}
[\mathcal{L}_1,\mathcal{L}_2]=\frac{[S_1,S_2^{\dagger}]-[S_1,S_2^{\dagger}]^{\dagger}}{4a_0^2}.
\label{SDPcomm}
\end{equation}
Finally, it remains to find the elements of $[S_1,S_2^{\dagger}]$. Both terms $S_1S_2^{\dagger}$ and $S_2^{\dagger}S_1$ produce the same net shift. The operator $S_1S_2^{\dagger}$ takes $\ket{k}\to\ket{k-k_2}\to\ket{k+k_1-k_2}$, while $S_2^{\dagger}S_1$ takes $\ket{k}\to\ket{k+k_1}\to\ket{k+k_1-k_2}$. An element of $[S_1,S_2^{\dagger}]$ thus survives only when one middle mode is blocked by the aperture while the other is transmitted. This can happen only near the edges of the aperture. Taking $k_1>k_2$ without loss of generality, the modes for which each ordering survives form an interval $\kc-k_1$, and the two intervals are displaced from one another by $k_2$. The two orderings thus clip differently over $\min\{k_2,\kc-k_1\}$ at each edge. As a fraction of $\kc$, we write this interval as
\begin{equation}
\mathcal{W}=\min\left\{\frac{k_2}{\kc},\;1-\frac{k_1}{\kc}\right\}
=\min_i\bigl\{\OTF(k_i),\,1-\OTF(k_i)\bigr\},
\label{SDPW}
\end{equation}
where we use $\OTF(k)=1-k/\kc$ for the one-dimensional aperture, and $\OTF(k_1)\le\OTF(k_2)$ to restore the symmetry between the two frequencies. Each term within $[S_1,S_2^{\dagger}]$ and $[S_1,S_2^{\dagger}]^{\dagger}$ contributes twice, giving
\begin{equation}
\TrAbs\bigl([S_1,S_2^{\dagger}]-[S_1,S_2^{\dagger}]^{\dagger}\bigr)
\ \le\ 4\,\mathcal{W}\,\frac{\kc L}{2\pi}.
\label{SDPcount}
\end{equation}
Substituting Eqs.~\eqref{SDPcomm} and \eqref{SDPcount} into Eq.~\eqref{SDP5}, we arrive at Eq.~\eqref{WZL_Eq6} of the main text:
\begin{equation}
\cNH\ \le\ \frac{2a_0^2}{\OTF(k_1)}+\frac{2a_0^2}{\OTF(k_2)}
+\frac{4a_0^2\,\mathcal{W}}{\OTF(k_1)\,\OTF(k_2)}
\ =\ \cQ+\frac{4a_0^2\,\mathcal{W}}{\OTF(k_1)\,\OTF(k_2)},
\label{SDP26}
\end{equation}
thereby upper-bounding Nagaoka's bound for our two-parameter estimation model. We plot this upper-bound in Figure~\ref{fig:WZL_Fig2.png} of the main text, without invoking approximation (ii) from Sec.~S3. The SDP finds the true NHCRB that lies below this derived bound. 

\section*{S5. DONN ARCHITECTURE AND TRAINING}\label{app:planes}
\noindent In this section, we specify the diffractive optical neural network's architecture, its training, the matched estimator, and the Monte Carlo verification protocol.


\medskip\noindent\textit{Architecture.}---Our receiver is a cascade of $P$ phase-only masks separated by optical Fourier transforms (realizable as $2f$ relays), followed by a photon-counting detector array in the output plane. The object field is sampled on $256$ points in one dimension and on a $64\times64$ grid in two; the propagation and detection grid spans twice the field of view, with $512$ and $128\times128$ samples respectively, and both the phase masks and the detector pixels are defined on this wider grid. Each mask $p$ applies the diagonal unitary $\exp[i\phi_p(\mathbf{r})]$ with trainable phases $\phi_p$, and each relay applies the unitary discrete Fourier transform.


\medskip\noindent\textit{Training.}---The training loss is the classical FIM \eqref{FI} of the photon-count distribution in the low-contrast limit: we minimize $\Tr[F^{-1}(\boldsymbol{\phi})]$ by gradient descent through the exact forward model, using the Adam optimizer with initial learning rates $0.04$--$0.05$ and retaining the best iterate
encountered along the trajectory. The masks are initialized with independent random phases of standard deviation $0.5$~rad; given the initialization the optimization is deterministic, and a single run per configuration proved sufficient. The single-parameter sweep of Fig.~\ref{fig:WZL_Fig1.png} uses 8 DONN layers at $4500$ epochs. The two-parameter sweep of Fig.~\ref{fig:WZL_Fig2.png} uses 10 planes at $6000$ epochs. The $M$-parameter points of Fig.~\ref{fig:WZL_Fig3.png} use $2M$ layers ($1500$ epochs). The two-dimensional networks of Fig.~\ref{fig:WZL_Fig4.png} use $128$ layers ($2500$ epochs). The networks of Figs.~\ref{fig:WZL_Fig3.png} and~\ref{fig:WZL_Fig4.png} are additionally polished with L-BFGS (strong Wolfe line search).

\medskip\noindent\textit{Matched estimator and Monte Carlo.}---
Because, as discussed in the main text, the detector counts are linear with respect to the mode amplitudes, estimating these amplitudes is a standard problem of linear statistics: the observed counts scatter around the linear model of Eq.~\eqref{Eq_DONN_probs}, and each output mode carries shot noise of variance approximately $Nu_d(0)$. The natural estimator is the weighted least-squares fit of the model to the data, with each mode weighted by the inverse of its noise variance $U=\mathrm{diag}\bigl[u_d(\boldsymbol{\theta})\bigr]$. This fit has a closed form, 
\begin{equation}
\hat{\boldsymbol{\theta}}
=F^{-1}B\,U^{-1}\!\left(\frac{\mathbf{n}}{N}-\mathbf{u}(0)\right).
\label{DT20}
\end{equation}
In this form, however, the estimator is object-dependent and must be applied iteratively. Therefore we make use of the low-contrast property of the object and simplify Eq.~\eqref{DT20} to the form quoted in Eq.~\eqref{WZL_Eq5} of the main text:
\begin{equation}
\hat{\boldsymbol{\theta}}
=F^{-1}B\,U_0^{-1}\!\left(\frac{\mathbf{n}}{N}-\mathbf{u}(0)\right),
\qquad U_0=\mathrm{diag}\bigl[u_d(0)\bigr],
\label{DT2}
\end{equation}
where $F=BU_0^{-1}B^{\top}$ is conveniently the FIM of Eq.~\eqref{FI}. In this way, the estimator becomes a linear map, fixed entirely by calibration data ($\mathbf{u}(0)$, $B$, and hence $F$), and it takes the full detection record to the full amplitude vector in a single measurement. This simplified estimator remains unbiased:
\begin{equation}
\mathbb{E}[\hat{\boldsymbol{\theta}}]
=F^{-1}BU_0^{-1}\bigl(\mathbf{u}(\boldsymbol{\theta})-\mathbf{u} (0)\bigr)
=F^{-1}BU_0^{-1}B^{\top}\boldsymbol{\theta}
=\boldsymbol{\theta}.
\label{DT3}
\end{equation}
Consider next the variance. For Poisson counting the frequencies are uncorrelated, with $\mathrm{Cov}[\mathbf{n}/N]=\mathrm{diag}[\mathbf{u}(\boldsymbol{\theta})]/N$, which at $\boldsymbol{\theta}=0$ equals $U_0/N$. Propagating this covariance through the linear map of Eq.~\eqref{DT2} gives
\begin{equation}
\mathrm{Cov}[\hat{\boldsymbol{\theta}}]
=F^{-1}BU_0^{-1}\,\frac{U_0}{N}\,U_0^{-1}B^{\top}F^{-1}
=\frac{F^{-1}}{N},
\label{DT4}
\end{equation}
the classical Cram\'er--Rao bound of the trained measurement, which the estimator attains. 

If the total photon number is instead held fixed, the counts are multinomial and $\mathrm{Cov}[\mathbf{n}/N]$ acquires the additional term $-\mathbf{u}\mathbf{u}^{\top}/N$; its contribution to Eq.~\eqref{DT4} is proportional to $BU_0^{-1}\mathbf{u}(0)$, which vanishes because $\sum_d B_{\mathbf{k}d}=\Tr[\partial_{\mathbf{k}}\rho]=0$. This leaves the conclusion unchanged. Finally, Eq.~\eqref{DT2} is also the one-step maximum-likelihood (Fisher-scoring) update started from $\boldsymbol{\theta}=0$: the score of the counting likelihood at $\boldsymbol{\theta}=0$ has components $\partial\ell/\partial a_{\mathbf{k}}=\sum_dn_dB_{\mathbf{k}d}/u_d(0)$, and multiplying it by $(NF)^{-1}$ reproduces Eq.~\eqref{DT2} by the same sum rule. The least-squares, maximum-likelihood, and Fisher-scoring methods thus give the same estimator.

In our simulations, we sample Poisson-distributed counts $n_d\sim\mathrm{Pois}[Nu_d(\boldsymbol{\theta})]$ for each pixel $d$ of the detector array, apply the estimator \eqref{DT2}, and record the empirical covariance of the estimates. For each dataset in Figs.~\ref{fig:WZL_Fig1.png}--\ref{fig:WZL_Fig4.png}, we perform $5000$ independent sampling runs, for which the expected relative scatter of the empirical variance is $\sqrt{2/5000}\approx2\%$. In all cases, the observed standard deviations match the Fisher predictions to within this Monte Carlo resolution. The direct-imaging Monte Carlo points are produced with only the DONN removed.

\medskip\noindent\textit{Calibration.}---The estimator \eqref{DT2} is assembled from $\mathbf{u}(0)$ and $B$. In simulation, both are calculated from the forward model. In an experiment, $\mathbf{u}(0)$ would be recorded with the object replaced by a structureless background of the same brightness, and $B$ is either computed from the characterized aperture and the programmed masks or measured directly, by presenting a known low-contrast reference pattern at each target frequency and recording the change in the output counts. Neither quantity involves the object of interest.


\end{document}